\documentclass{INTERSPEECH2023}
\usepackage[capitalize,noabbrev]{cleveref}
\usepackage{booktabs}
\usepackage[table,x11names]{xcolor}
\newcolumntype{H}{>{\setbox0=\hbox\bgroup}c<{\egroup}@{}}

\usepackage{hyperref}

 \interspeechcameraready

\title{UnDiff: Unsupervised Voice Restoration with Unconditional Diffusion Model}
\name{Anastasiia Iashchenko$^{123*}$, Pavel Andreev$^{1*}$, Ivan Shchekotov$^{123*}$, Nicholas Babaev$^{1}$, Dmitry Vetrov$^{24}$}
\address{
  $^1$Samsung Research\quad $^2$HSE University, Moscow \\
  $^3$Skolkovo Institute of Science and Technology, Moscow \\
  $^4$Artificial Intelligence
Research Institute, Moscow \quad $^*$equal contribution}
\email{p.andreev@samsung.com, a.yaschenko@partner.samsung.com, i.shchekotov@samsung.com}
\begin{document}

\maketitle
 
\begin{abstract}
This paper introduces UnDiff, a diffusion probabilistic model capable of solving various speech inverse tasks. Being once trained for speech waveform generation in an unconditional manner, it can be adapted to different tasks including degradation inversion, neural vocoding, and source separation. In this paper, we, first, tackle the challenging problem of unconditional waveform generation by comparing different neural architectures and preconditioning domains. After that, we demonstrate how the trained unconditional diffusion could be adapted to different tasks of speech processing by the means of recent developments in post-training conditioning of diffusion models. Finally, we demonstrate the performance of the proposed technique on the tasks of bandwidth extension, declipping, vocoding, and speech source separation and compare it to the baselines. The codes are publicly available\footnote{\href{https://github.com/SamsungLabs/Undiff}{https://github.com/SamsungLabs/Undiff}}.
\end{abstract}

\section{Introduction}

As the field of artificial intelligence continues to evolve, generative models have emerged as a powerful tool for a variety of tasks including speech processing. In recent years, diffusion models~\cite{song2020score,ho2020denoising,karras2022elucidating} have gained attention due to their ability to efficiently model complex high-dimensional distributions. Diffusion models are designed to learn the underlying data distribution's implicit prior by matching the gradient of the log density. This learned prior can be useful for solving inverse problems, where the objective is to recover the input signal $x$ from the measurements $y$, which are typically linked through some differentiable operator $A$, s.t. $y = A(x) + n$, where $n$ is some noise. In this paper, we introduce UnDiff, a diffusion probabilistic model specifically designed to tackle various inverse tasks for speech processing including degradation inversion, neural vocoding, and source separation.

The key advantage of UnDiff is its ability to be trained in an unconditional manner for speech waveform generation and then be adapted for the inverse problem without any additional supervised training. This is in contrast to existing approaches that utilize conditional diffusion models for waveform restoration and generation or design specific training pipelines for specific tasks~\cite{serra2022universal,richter2022speech, scheibler2022diffusion}. Similarly to our work recent paper~\cite{moliner2022solving} utilizes an unconditional diffusion model for piano music restoration solving declipping, bandwidth extension, and inpainting problems. Unlike this work, we tackle a more challenging problem of speech restoration and additionally consider neural vocoding and speech source separation problems which we formulate as inverse problems.  

We explore the challenges of unconditional waveform generation and compare different neural architectures and preconditioning domains. Furthermore, we demonstrate the effectiveness of UnDiff in solving a variety of speech processing tasks such as bandwidth extension, declipping, neural vocoding, and speech source separation.  We utilize recent developments in diffusion guided sampling~\cite{chung2022diffusion, song2020score, choi2021ilvr} to adapt the unconditional diffusion for each task. Remarkably, this work proposes a novel diffusion inverse task solver for source separation showing tractability of log-likelihood for this case.

The results show that UnDiff performs comparably with baselines, making it a promising solution for a variety of speech processing tasks. Overall, this paper highlights the potential of diffusion models in solving general inverse problems for speech processing and provides a new direction for future research in this field.

\section{Background}

\subsection{Score-based diffusion models}

Score-based diffusion models~\cite{song2020score} are the class of neural generative models, that can be informally described as gradually transforming analytically known and unknown (only samples are available) data distributions $p_{\text{known}}$ and $p_{\text{data}}$ to each other. More formally, one can consider a forward (\ref{forward_sde}) and reverse (\ref{reverse_sde}) Ito stochastic equations (VP-SDE) for the data noising process in the following form
\begin{equation}
    d \mathbf{x} = - \frac{\beta(t)}{2} \mathbf{x} dt + \sqrt{\beta(t)} d \textbf{w}, \label{forward_sde}
\end{equation}
\begin{equation}
    d \mathbf{x} = \bigg (- \frac{\beta(t)}{2} \mathbf{x} - \beta(t) \nabla_{\mathbf{x}_t} \log p_t(\mathbf{x}_t)  \bigg )dt + \sqrt{\beta(t)} d \textbf{w},  \label{reverse_sde}
\end{equation}
where  $t \in [0, T]$ is the time variable, $\beta(t)$ is the noise schedule of the process, chosen such that if $\mathbf{x}_0 \sim p_{\text{data}}$, then $\mathbf{x}_T \sim p_{\text{known}} = \mathcal{N}(\mathbf{0}, \mathbf{I})$, $\mathbf{w}$ is Wiener process. While other forms of stochastic differential equations exist in the literature, throughout this paper we employ VP-SDE, which is equivalent to DDPM~\cite{ho2020denoising}. 

 Once the score function $\nabla_{\mathbf{x}_t}\log p_t(\mathbf{x}_t)$ is known, it is possible to solve reverse SDE (\ref{reverse_sde}) numerically and thus generate samples from $p_{\text{data}}$.  It can be shown that the score function could be approximated by a neural network $\mathbf{s}_{\theta}(\mathbf{x}_t, t)$ trained with denoising score matching objective eventually leading to L2 loss function:

\begin{equation}
\mathbb{E}_{\mathbf{x} \sim p_{\text{data}}, \varepsilon \sim \mathcal{N}(\mathbf{0}, \mathbf{I})} \bigg ( \lambda(t) \big \lVert \mathbf{s}_\theta (\mathbf{x}_t, t) - \frac{\mathbf{\varepsilon}}{\sqrt{1 -  \bar{\alpha}(t)}} \big \rVert_2^2 \bigg ), 
\end{equation}
where $\lambda(t)$ is some weighting function, $\mathbf{x}_t = \sqrt{ \bar{\alpha}(t)}\mathbf{x}_0 + \sqrt{1 -  \bar{\alpha}(t)} \mathbf{\varepsilon}$,  and  $ \bar{\alpha}(t)$ is an  explicit function of $\beta(t)$. In practice we optimize scaled version of score $\varepsilon_\theta (\mathbf{x}_t, t) =  \mathbf{s}_\theta (\mathbf{x}_t, t) \cdot \sqrt{1- \bar{\alpha}(t)}$ and set $\lambda(t) = 1$. We use  linear schedule for $\beta(t) \in [0.0001, 0.02]$ and set $ \bar{\alpha}(t) = \prod_{s=1}^{t}(1-\beta(s))$.

\subsection{Inverse problems with diffusion models}
The inverse problems address the task of retrieving object $\mathbf{x}$ given its partial observation $\mathbf{y}$ and the degradation model $p(\mathbf{y} | \mathbf{x})$. To utilize reverse SDE (\ref{reverse_sde}) for sampling from conditional distribution $p(\mathbf{x} | \mathbf{y})$, one needs to know the score function of conditional distribution $\nabla_{\mathbf{x}_t} \log p_t(\mathbf{x}_t | \mathbf{y})$. 

One way to estimate $\nabla_{\mathbf{x}_t} \log p_t(\mathbf{x}_t | \mathbf{y})$ is to apply imputation guidance (data consistency) \cite{song2020score, moliner2022solving, choi2021ilvr}. The idea of this method is to explicitly modify the score so that some parts of a denoised estimate $ \mathbf{\hat x}_0 = \frac{1}{\sqrt{ \bar{\alpha}(t)}} (\mathbf{x}_t - (1 -  \bar{\alpha}(t)) s_\theta(\mathbf{x}_t, t))$ are imputed with observations $\mathbf{y}$. We later elaborate on how imputation could be used for our problem.

Another way to formalize the search for $\mathbf{x}$ is the usage of Bayes’ rule:
\begin{equation}
    p(\mathbf{x}|\mathbf{y}) = p(\mathbf{y} | \mathbf{x}) p(\mathbf{x}) / p(\mathbf{y}),
\end{equation}
thus,
\begin{equation}
   \nabla_{\mathbf{x}_t} \log p_t(\mathbf{x}_t|\mathbf{y}) = \nabla_{\mathbf{x}_t}  \log p_t(\mathbf{y} | \mathbf{x}_t) + \nabla_{\mathbf{x}_t}  \log p_t(\mathbf{x}_t), \label{br}
\end{equation}
 $\nabla_{\mathbf{x}_t}  \log p_t(\mathbf{y} | \mathbf{x}_t)$ is generally intractable. However, \cite{chung2022diffusion} showed that one can make approximation  $\nabla_{\mathbf{x}_t}  \log p(\mathbf{y} | \mathbf{x}_t) \approx \nabla_{\mathbf{x}_t}  \log p(\mathbf{y} |  \mathbf{\hat x}_0)$, where $ \nabla_{\mathbf{x}_t} \log p(\mathbf{y} |  \mathbf{\hat x}_0)$ can be computed using degradation model. Given observation operator $A$ and assuming Gaussian likelihood, the final approximation becomes:
  \begin{equation}
      \nabla_{\mathbf{x}_t}  \log p_t(\mathbf{y} | \mathbf{x}_t) \approx -\xi (t) \nabla_{\textbf{x}_t} \lVert \mathbf{y} - A(\mathbf{\hat x}_0)\rVert^2_2
  \end{equation}
where $\xi (t)$ weighting coefficient which we set to be inversely proportional to the gradient norm similarly to \cite{moliner2022solving}.
 Likewise \cite{moliner2022solving} we refer to this method as reconstruction guidance.

\section{UnDiff}

\subsection{Unconditional speech generation}

Unconditional speech generation is a challenging task due to the high diversity of possible linguistic content. 
Prior works on diffusion models tend to consider conditional speech generation~\cite{kong2020diffwave, liu2023audioldm} or limit the scope to simple datasets with predefined phrases (e.g., spoken digits)~\cite{goel2022s, kong2020diffwave}. 
Unlike these works, we aim to train the unconditional diffusion model and do not constrain the linguistic content of the datasets.  We consider three approaches to building unconditional diffusion models, all approaches operate in the time domain but have different preconditioning transformations:
\begin{enumerate}
    \item Diffwave~\cite{kong2020diffwave} neural network operating directly in time domain;
    \item FFC-AE~\cite{shchekotov22_interspeech} neural network operating on short-time Fourier transform spectrograms;
    \item UNet~\cite{moliner2022solving} neural network operating  on Constant-Q transform spectrograms.
\end{enumerate}

\subsection{Speech inverse tasks}

\noindent \textbf{Bandwidth extension} \ \ \  Frequency bandwidth extension \cite{kuleshov2017audio, andreev2022hifi++} (also known as audio super-resolution) can be viewed as a realistic restoration of waveform's high frequencies. The observation operator is a lowpass filter $\mathbf{y} = A(\mathbf{x}) = \mathrm{LPF}(\mathbf{x})$. Thus, imputation guidance in this case corresponds to substituting the generated estimate of low frequencies with observed low frequencies $\mathbf{y}$ at each step. More formally, this corresponds to modifying the score function during sampling as 
 \begin{equation}
    \Tilde{s}_\theta(\mathbf{x}_t, t) = \frac{1}{1 -  \bar{\alpha}(t)} (\sqrt{ \bar{\alpha}(t)} \mathbf{\Tilde{\hat x}}_0 - \mathbf{x}_t),
 \end{equation}
 
 where $ \mathbf{\Tilde{\hat x}}_0 = \mathbf{\hat x}_0 - \mathrm{LPF}(\mathbf{\hat x}_0) + \mathbf{y}$ is imputed estimate of $\mathbf{x}_0$, and $\mathbf{\hat x}_0 = \frac{1}{\sqrt{ \bar{\alpha}(t)}} (\mathbf{x}_t + (1 -  \bar{\alpha}(t)) s_\theta(\mathbf{x}_t, t))$ is estimate of $\mathbf{x}_0$ with original score function.

\noindent \textbf{Declipping} \ \ \  Similarly to \cite{moliner2022solving} we consider clipping as an inverse problem with observation function defined as $A = \mathrm{clip}(\mathbf{x}) = \frac{1}{2} ( | x + c  | - | x - c  |)$ and apply reconstruction guidance strategy. 


\noindent \textbf{Neural vocoding} \ \ \  The majority of modern speech synthesis systems decompose this task into two stages. 
In the first stage, low-resolution intermediate representations (e.g., linguistic features, mel-spectrograms) are predicted from text data \cite{li2020robutrans, shen2020non}. 
In the second stage, these intermediate representations are transformed to raw waveform \cite{kong2020hifi,prenger2019waveglow}.
Neural vocoders relate to the techniques used in the second stage of the speech synthesis process.
The neural vocoding can be formulated as the inverse problem with the observation operator defined as mel-spectrogram computation $A(\mathbf{x}) = \mathrm{Mel}(\mathbf{x})$. Since mel-spectrogram computation is a differentiable operation we can easily apply reconstruction guidance in this case.


\noindent \textbf{Source separation} \ \ \  The goal of single-channel speech separation is to extract individual speech signals from a mixed audio signal, in which multiple speakers are talking simultaneously. The potential applications of speech source separation include teleconferencing, speech recognition, and hearing aid technology. Let $\mathbf{x}_1$ and $\mathbf{x}_2$ be the two voice recordings. Consider the observation model which mixes these two recordings, i.e., $\mathbf{y} = A(\mathbf{x}_1, \mathbf{x}_2) =  \mathbf{x}_1 +  \mathbf{x}_2$. Note that since $\mathbf{x}_1$ and $\mathbf{x}_2$ are independent, unconditional density function on their joint distribution can be factorized as $p(\mathbf{x}_1, \mathbf{x}_2) = p(\mathbf{x}_1) \cdot p(\mathbf{x}_2)$. Thus, for the unconditional score function of joint distribution, we have
\begin{multline}
         \nabla_{ [ \mathbf{x}_{1, t},  \mathbf{x}_{2, t}] }  \log p_t (\mathbf{x}_{1, t},  \mathbf{x}_{2, t}) =  
        [ \nabla_{ \mathbf{x}_{1, t} }  \log p_t (\mathbf{x}_{1, t}), \mathbf{0} ] + \\
        [\mathbf{0}, \nabla_{ \mathbf{x}_{2, t}}  \log p_t (\mathbf{x}_{2, t}) ], \label{uncond_ll}
\end{multline}
According to (\ref{br}), to sample from joint conditional density, we need to also estimate the gradient of log-likelihood $ \nabla_{ [ \mathbf{x}_{1, t},  \mathbf{x}_{2, t}]}  \log p_t(\mathbf{y} | \mathbf{x}_{1, t},  \mathbf{x}_{2, t})$. One can apply reconstruction guidance (6), however, we found a more natural way to estimate the log-likelihood gradient in this case. Specifically, note that $\mathbf{y}$ depends only the sum of $\mathbf{x}_1$ and $\mathbf{x}_2$, it can be shown that the same holds for  $\mathbf{x}_{1, t}$ and $\mathbf{x}_{2, t}$, i.e.,  $  p_t(\mathbf{y} | \mathbf{x}_{1, t},  \mathbf{x}_{2, t}) = p_t(\mathbf{y} | \mathbf{x}_{1, t} +  \mathbf{x}_{2, t})$. This likelihood can be computed analytically, indeed, since $\mathbf{x}_{1, t} = \sqrt{ \bar{\alpha}(t)}\mathbf{x}_{1, 0} + \sqrt{1 -  \bar{\alpha}(t)} \mathbf{\varepsilon}_1$ and $\mathbf{x}_{2, t} = \sqrt{ \bar{\alpha}(t)}\mathbf{x}_{2, 0} + \sqrt{1 -  \bar{\alpha}(t)} \mathbf{\varepsilon}_2$, where $\varepsilon_1, \varepsilon_2  \sim \mathcal{N}(\mathbf{0}, \mathbf{I})$, we have
\begin{equation}
   \mathbf{y} =  \mathbf{x}_{1} + \mathbf{x}_{2} = \frac{1}{\sqrt{ \bar{\alpha}(t)}}(\mathbf{x}_{1, t} + \mathbf{x}_{2, t}) - \sqrt{\frac{1 -  \bar{\alpha}(t)}{ \bar{\alpha}(t)}} (\varepsilon_1 + \varepsilon_2)
\end{equation}
 Thus, $p_t(\mathbf{y} | \mathbf{x}_{1, t} +  \mathbf{x}_{2, t}) = \mathcal{N}(\mathbf{y}; \frac{1}{\sqrt{ \bar{\alpha}(t)}}(\mathbf{x}_{1, t} + \mathbf{x}_{2, t}), 2 \cdot \frac{1 -  \bar{\alpha}(t)}{ \bar{\alpha}(t)})$. Therefore, we can compute the gradient of the log-likelihood analytically 
 \begin{equation}
      \nabla_{  \mathbf{x}_{1, t}}  \log p_t(\mathbf{y} | \mathbf{x}_{1, t},  \mathbf{x}_{2, t}) = \frac{\sqrt{ \bar{\alpha}(t)} (\mathbf{y} - \frac{1}{\sqrt{ \bar{\alpha}(t)}} (\mathbf{x}_{1, t} + \mathbf{x}_{2, t}))}{2 (1 -  \bar{\alpha}(t))}, \label{cond_ll}
 \end{equation}
 the same relation holds for $ \nabla_{  \mathbf{x}_{2, t}}  \log p_t(\mathbf{y} | \mathbf{x}_{1, t},  \mathbf{x}_{2, t})$.
 


\section{Experiments and discussion}
\subsection{Datasets}
We use two datasets for our experiments. 

The first one is the publicly available VCTK dataset~\cite{yamagishi2019cstr} which includes 44200 speech recordings belonging to 110 speakers. We exclude 6 speakers from the training set and 8 recordings from the utterances corresponding to each speaker to avoid text-level and speaker-level data leakage to the training set. For evaluation, we use 48 utterances corresponding to 6 speakers excluded from the training data. 

The second dataset is  the LJ-Speech dataset \cite{ljspeech17} which is standard in the speech synthesis field. LJ-Speech is a single-speaker dataset that consists of 13,100 audio clips with a total length of approximately 24 hours. We use train-validation split from \cite{kong2020hifi} with sizes of 12950 train clips and 150 validation clips. Audio samples have a sampling rate of 22.05 kHz.
\subsection{Metrics}
 For the evaluation of samples generated by the unconditional model, we use publicly available absolute objective speech quality measure based on direct MOS score prediction by a fine-tuned wav2vec2.0 \cite{NEURIPS2020_wav2vec2} model (WV-MOS~\cite{wvmos}) and unconditional Frechet DeepSpeech Distance (FDSD) introduced in \cite{binkowski2019high}. WV-MOS measures the quality of each generated sample individually while FDSD measures the distance between distributions of generated and real samples. For quality evaluation in speech inverse tasks, we use conventional metrics extended STOI \cite{jensen2016algorithm}, scale-invariant signal-to-noise ratio (SI-SNR) \cite{le2019sdr}, log-spectral distance (LSD)
 and WV-MOS.  We also use 5-scale MOS tests for subjective quality evaluation following the procedure described in \cite{andreev2022hifi++}. 

\subsection{Experimental details}
We train all models for 230 epochs. The models are trained with batch size 8 on audio segments of 2 seconds at the sampling rate of 16 kHz. We use Adam optimizer with a learning rate of 0.0002 and betas 0.9 and 0.999. For DDPM training, we perform denoising over 200 steps during training, and condition on  $\beta(t)$. All models were trained for approximately 6 days on 4 A100 GPUs. 

\subsection{Unconditional speech generation}
We compare 3 approaches to unconditional diffusion-based speech generation and 3 additional baseline cases. All considered approaches operate in the time domain but use different invertible preprocessing transformations and their corresponding inverse postprocessing transformations for the preconditioning of neural networks. To select the best architecture, we tune hyperparameters of all models so that they have equal capacities as measured by GPU memory allocated for training each model with equal batch size (completely utilizing the capacity of 4 A100 GPUs).

The first approach is the training of neural architecture directly in the time domain, i.e., without any preconditioning transformation. After our preliminary experiments, we found that Diffwave architecture provides the best performance among tested time-domain architectures (we also tried UNIVERSE~\cite{serra2022universal} and UNet~\cite{nichol2021improved}). 
We redistribute the capacity of the original unconditional Diffwave architecture by increasing the number of blocks from 36 to 48 and keeping the number of channels equal to 256. Additionally, we introduce squeeze-excitation \cite{hu2018senet} weighting on skip connections, and condition generative model on $\beta(t)$ via random Fourier features \cite{crash21}. We found that these modifications significantly improved the performance compared to the original Diffwave architecture.

Another approach is based on time-frequency domain architecture FFC-AE~\cite{shchekotov22_interspeech} which uses short-time Fourier transform (STFT) as preconditioning. This architecture is based on a fast Fourier convolution neural operator and operates on complex-valued STFT spectrograms. We found FFC-AE to provide superior quality compared to convolutional UNet-type architectures. We use FFC-AE architecture consisting of 18 blocks with 256 channels, with conditioning done similarly to Diffwave.

Finally, we test the approach to unconditional audio generation as proposed in \cite{moliner2022solving}. In this approach, we use Constant-Q Transform (CQT) as a preconditioning transformation and convolutional UNet neural architecture with dilated residual blocks as a neural architecture as recommended by~\cite{moliner2022solving}. We use UNet of depth 5 with the following channels =  [64, 64, 128, 128, 256], with downsampling by a factor of 2 at each block. 

We compare the quality of 8000 unconditionally generated samples based on WV-MOS and FDSD metrics. We also provide metrics for 4 baseline cases: ground-truth speech, gaussian noise, samples from unconditional Diffwave with original architecture~\cite{kong2020diffwave},  and text-to-audio AudioLDM~\cite{liu2023audioldm}  model generated with the prompt "A person speaking English". The results are presented in Table \ref{table:uncond}.

\begin{table}[!h]
  \vspace{-0.1cm}
    \centering
    \caption{Results of unconditional speech generation (VCTK).}
      \vspace{-0.2cm}
	\label{tab:phase}
 \scalebox{0.9}{
	\begin{tabular}{l c c  | c H}
		\toprule
		Model & WV-MOS (\textuparrow
) &  FDSD (\textdownarrow
) & \# Params (M) & \# GMAC on 8k \\
		\midrule
		\midrule
		Ground Truth & 4.57  & 0.9 & - & - \\
		\midrule
		FFC-AE  & $\mathbf{4.06 }$  & 15.3 & 55.3 & - \\
		Diffwave (ours) & 3.84  &  \textbf{7.0}& 32.3& - \\
		  CQT-UNet & 2.29 &  12.37& 27.8 & - \\
            \midrule
            AudioLDM & 1.81   & 22.5 & 185.0 & - \\
            Diffwave (orig.) &  3.12 & \textbf{7.0} & 24.0 & - \\
            Gaussian noise & 1.27  & 153.5 & - & - \\
			\bottomrule
	\end{tabular} }\label{table:uncond}
	 \vspace{-0.2cm}
\end{table}


Overall, all the models demonstrate the ability to generate speech-like sounds but do not produce any semantically consistent speech. This behavior is rather expected since we do not constrain the linguistic content of the training dataset and do not provide any language understanding guidance (unlike, e.g., AudioLM \cite{borsos2022audiolm}). However, we believe that language understanding is not necessary for speech restoration since voice could be potentially retrieved based on acoustic (syntactic) information.  We provide examples of sounds generated by our model in the supplementary material.


We observed that the FFC-AE model provides better WV-MOS quality, while Diffwave delivers the lowest FDSD score. Since it is not clear what property is more important for solving downstream inverse tasks, we conduct our subsequent experiments with both Diffwave and FFC-AE models. 

\subsection{Inverse tasks}

The experimental results for bandwidth extension, declipping, neural vocoding, and source separation are provided in Tables 2, 3, 4, and 5 (best results are highlighted in bold). For qualitative evaluation, we attach examples of Undiff outputs for each inverse task as a part of supplementary material. All the metrics were computed on randomly cropped 1-second segments.  

\noindent \textbf{Bandwidth extension} \ \ \  In our bandwidth extension experiments, we use recordings
with a sampling rate of 16 kHz as targets and consider two frequency bandwidths for input data: 2 kHz and 4
kHz. We artificially degrade the signal to the desired frequency bandwidth (2 kHz or 4 kHz) using polyphase filtering. 
The results and comparison with other techniques are outlined in Table \ref{table:bwe}.

\begin{table}[!h]
    \centering
    \caption{Results of bandwidth extension (BWE) on VCTK.}
      \vspace{-0.2cm}
	\scalebox{0.9}{
 \begin{tabular}{l c c c   H c}
		\toprule
		Model & Supervised  & WV-MOS &  LSD & ESTOI & MOS \\
		\midrule
		\midrule
		Ground Truth & - & 4.17 & 0 & 1.0 & $4.09 \pm 0.09$ \\
            \midrule
            \multicolumn{2}{c}{\textbf{BWE 2kHz} $\xrightarrow{}$\textbf{8kHz}} && \\
            \midrule
            HiFi++~\cite{andreev2022hifi++}  & $\checkmark$ & \textbf{4.05} & 1.09 & 0.59 & $\mathbf{3.93  \pm  0.10}$ \\
            Voicefixer~\cite{liu2021voicefixer} & $\checkmark$ & 3.67 & 1.08 & 0.62 & $3.64 \pm 0.10 $ \\
            TFiLM~\cite{birnbaum2019temporal}  & $\checkmark$& 2.83 & 1.01 & \textbf{0.81} & $2.71 \pm 0.10$ \\
              \midrule
            UnDiff (Diffwave) & $\times$ & 3.48 & \textbf{0.96} & 0.80 & $3.59 \pm 0.11$ \\
            UnDiff (FFC-AE) & $\times$  & 3.59 & 1.13 & 0.76 & $3.50 \pm 0.11$ \\
  \midrule
            Input  & - & 2.52 & 1.06 & 0.66 &  $2.42 \pm 0.09 $ \\
            \midrule
            \multicolumn{2}{c}{\textbf{BWE 4kHz} $\xrightarrow{}$\textbf{8kHz}} &&\\
            \midrule
            HiFi++~\cite{andreev2022hifi++}  & $\checkmark$ & \textbf{4.22} & 1.07 & 0.65 & $\mathbf{4.04 \pm 0.10} $\\
            Voicefixer~\cite{liu2021voicefixer}   & $\checkmark$ & 3.95 & 0.98 & 0.73 & $3.92 \pm 0.10$\\
            TFiLM~\cite{birnbaum2019temporal}   & $\checkmark$ & 3.46 & 0.83 & \textbf{0.99} & $3.43 \pm 0.10$\\
              \midrule
            UnDiff (Diffwave)  & $\times$ & 4.00 & \textbf{0.76} & 0.98 & $3.74 \pm 0.11 $ \\
            UnDiff (FFC-AE)  & $\times$ & 3.88 & 0.96 & 0.97 & $3.72 \pm 0.10$ \\
              \midrule
            Input   & - & 3.34 & 0.85 & 0.99 & $3.39 \pm 0.10$ \\
			\bottomrule
	\end{tabular}}\label{table:bwe} 
	 \vspace{-0.2cm}
\end{table}


\noindent \textbf{Decliping} \ \ \    We compare our models against popular audio declipping methods A-SPADE~\cite{zavivska2022analysis} and S-SPADE~\cite{zavivska2019proper}, as well as the general speech restoration framework Voicefixer~\cite{liu2021voicefixer} on clipped audio recordings with input SDR being equal to 3 db (see Table 3). 

\begin{table}[!h]
  \vspace{-0.2cm}
    \centering
    \caption{Results of declipping (input SNR $=$ 3 db) on VCTK.}
      \vspace{-0.2cm}
	\label{tab:phase}
 \scalebox{0.9}{
	\begin{tabular}{l c c H   c c}
		\toprule
		Model & Supervised & WV-MOS &  PESQ &  SI-SNR & MOS \\
		\midrule
		\midrule
  		Ground Truth & - & 3.91  & 4.33 & - & $3.84 \pm 0.11$ \\
  \midrule

  

     A-SPADE~\cite{zavivska2022analysis} & $\times$ & 2.63  & 1.58 & 8.48 & $2.67 \pm 0.11 $ \\
     S-SPADE~\cite{zavivska2019proper} & $\times$ & 2.69  & 1.60 & 8.50 & $2.55 \pm 0.11$ \\
     Voicefixer~\cite{liu2021voicefixer}& $\checkmark$ & 2.79  & 1.28 & -22.58 & $2.98 \pm 0.12$ \\
       \midrule
	 Undiff (Diffwave) & $\times$& \textbf{3.62} & \textbf{1.88} & \textbf{10.57} & $\mathbf{3.59 \pm 0.12}$ \\
   	 Undiff (FFC-AE) & $\times $ & 3.01 & 1.57 & 7.35 & $3.06 \pm 0.12$ \\
       \midrule
  	Input  & - & 2.30  & 1.28 & 3.82 & $2.19 \pm 0.09$ \\
			\bottomrule
	\end{tabular} 
 }\label{table:clip}
	 \vspace{-0.2cm}
\end{table}
\noindent \textbf{Neural vocoding} \ \ \  To demonstrate the effectiveness of the Undiff model on neural vocoding, we train FFC-AE and Diffwave models on the unconditional generation of the LJ-speech dataset. We compare our approach with 2 supervised baselines from the literature and the unsupervised Griffin-Lim vocoder.  
\begin{table}[!h]
  \vspace{-0.2cm}
    \centering
    \caption{Results of neural vocoding (LJ speech dataset).}
      \vspace{-0.1cm}
	\label{tab:phase}
	\begin{tabular}{l c c   c H}
		\toprule
		Model & Supervised &  WV-MOS &  MOS & \# GMAC on 8k \\
		\midrule
		\midrule
		Ground Truth & - & 4.32 & $4.26 \pm 0.07$ & - \\
  \midrule
            HiFi-GAN (V1) ~\cite{kong2020hifi} &  $\checkmark$ & \textbf{4.36} & $\mathbf{4.23 \pm 0.07}$ & - \\
            Diffwave~\cite{kong2020diffwave} & $\checkmark$ & 4.19 & $4.15    \pm 0.07$ & - \\
            Griffin-Lim~\cite{griffin1984signal} &  $\times$ & 3.30 & $3.46    \pm 0.08 $ & - \\
              \midrule
            Undiff (Diffwave) &  $\times$ & 3.99 & $3.79 \pm 0.08$ & - \\
            Undiff (FFC-AE) &  $\times$ & 4.08 & $4.12 \pm 0.07$ & - \\

			\bottomrule
	\end{tabular} \label{table:vocod}
	 \vspace{-0.3cm}
\end{table}

\begin{figure}[!h]
  \centering
  \includegraphics[width=0.95\linewidth]{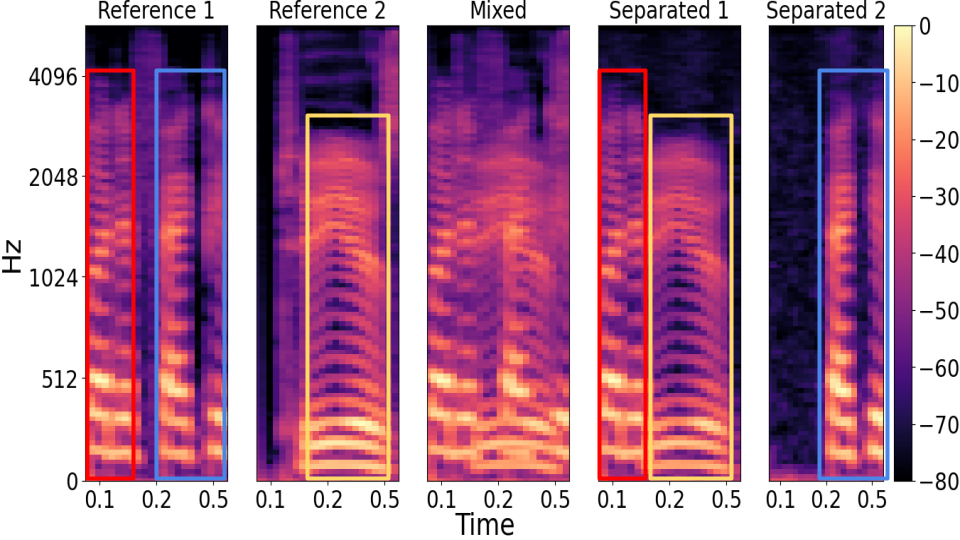}
  \caption{Failure case of source separation with Undiff model. }
  \label{fig:failure}
  \vspace{-0.1cm}
\end{figure}

\noindent \textbf{Source separation} \ \ \  To assess the Undiff's performance on the source separation task, we randomly mix recordings belonging to different speakers from VCTK validation data. The recordings were normalized and mixed without a weighting coefficient. Though being far from real-life, such mixing makes the source separation task to be easier for the unsupervised diffusion model and allows for identifying pitfalls in unsupervised speech source separation. While Undiff is capable to solve this task to some extent, we found that even in such a scenario it performs significantly worse than supervised baseline ConvTasNet. One of the most characteristic artifacts that we observe is the inability of the model to correctly identify the global context. Although Undiff is able to correctly separate voices in local regions, it mixes different voices within one sample (Figure \ref{fig:failure}).

\begin{table}[!h]
  \vspace{-0.1cm}
    \centering
    \caption{Results of source separation (VCTK dataset).}
      \vspace{-0.2cm}
	\label{tab:phase}
  \scalebox{0.9}{
	\begin{tabular}{l c H c   c H}
		\toprule
		Model & Supervised& WV-MOS &  SI-SNR &  STOI & \# GMAC on 8k \\
		\midrule
		\midrule
        Mixture (input) & - & 3.10 $\pm 0.28 $ &-0.04 &  0.69 & - \\
        \midrule
		Undiff (Diffwave) & $\times$ & 2.59 $\pm$ 0.48 &5.73 &  0.79 & - \\
  		Undiff (FFC-AE) & $\times$ & 2.25 $\pm$ 0.69 & 3.39 &  0.76 & - \\
      \midrule
        Conv-TasNet~\cite{luo2019conv} & $\checkmark$  & 4.02 $\pm$ 0.26 & \textbf{15.94} &  \textbf{0.95} & - \\

			\bottomrule
	\end{tabular}} \label{table:sep}
	 \vspace{-0.5cm}
\end{table}

The results show that despite the Undiff was never explicitly trained to solve any of the considered tasks, it performs comparably to supervised baselines for bandwidth extension, declipping and vocoding. It also demonstrates the potential to solve the source separation task, although there are still some significant challenges to overcome. An interesting directions for future work could be considering different mixing weights and enabling models to produce globally coherent voices during source separation. Overall, the results highlight the potential of the unconditional diffusion models to serve as general voice restoration tools. 
  \vspace{-0.2cm}
\section{Conclusion}

In this paper, we introduced UnDiff, a diffusion probabilistic model capable of solving various speech inverse tasks. We demonstrated the performance of the model in bandwidth extension, declipping, neural vocoding, and source separation tasks. The development of UnDiff provides a new tool for solving complex inverse problems in speech restoration, highlighting the potential of diffusion models to be a general framework for voice restoration.
\vspace{-0.2cm}
\section{Acknowledgements}
This work was supported by Samsung Research. Dmitry Vetrov
was supported by the grant provided by the Analytical Center
for the Government of the Russian Federation (ACRF) in accor-
dance with the agreement No. 000000D730321P5Q0002 and
the agreement with HSE University No. 70-2021-00139.

\bibliographystyle{IEEEtran}
\bibliography{A}

\begin{thebibliography}{10}
\providecommand{\url}[1]{#1}
\csname url@samestyle\endcsname
\providecommand{\newblock}{\relax}
\providecommand{\bibinfo}[2]{#2}
\providecommand{\BIBentrySTDinterwordspacing}{\spaceskip=0pt\relax}
\providecommand{\BIBentryALTinterwordstretchfactor}{4}
\providecommand{\BIBentryALTinterwordspacing}{\spaceskip=\fontdimen2\font plus
\BIBentryALTinterwordstretchfactor\fontdimen3\font minus
  \fontdimen4\font\relax}
\providecommand{\BIBforeignlanguage}[2]{{%
\expandafter\ifx\csname l@#1\endcsname\relax
\typeout{** WARNING: IEEEtran.bst: No hyphenation pattern has been}%
\typeout{** loaded for the language `#1'. Using the pattern for}%
\typeout{** the default language instead.}%
\else
\language=\csname l@#1\endcsname
\fi
#2}}
\providecommand{\BIBdecl}{\relax}
\BIBdecl

\bibitem{song2020score}
Y.~Song, J.~Sohl-Dickstein, D.~P. Kingma, A.~Kumar, S.~Ermon, and B.~Poole,
  ``Score-based generative modeling through stochastic differential
  equations,'' in \emph{International Conference on Learning Representations}.

\bibitem{ho2020denoising}
J.~Ho, A.~Jain, and P.~Abbeel, ``Denoising diffusion probabilistic models,''
  \emph{Advances in Neural Information Processing Systems}, vol.~33, pp.
  6840--6851, 2020.

\bibitem{karras2022elucidating}
T.~Karras, M.~Aittala, T.~Aila, and S.~Laine, ``Elucidating the design space of
  diffusion-based generative models,'' in \emph{Advances in Neural Information
  Processing Systems}.

\bibitem{serra2022universal}
J.~Serr{\`a}, S.~Pascual, J.~Pons, R.~O. Araz, and D.~Scaini, ``Universal
  speech enhancement with score-based diffusion,'' \emph{arXiv preprint
  arXiv:2206.03065}, 2022.

\bibitem{richter2022speech}
J.~Richter, S.~Welker, J.-M. Lemercier, B.~Lay, and T.~Gerkmann, ``Speech
  enhancement and dereverberation with diffusion-based generative models,''
  \emph{arXiv preprint arXiv:2208.05830}, 2022.

\bibitem{scheibler2022diffusion}
R.~Scheibler, Y.~Ji, S.-W. Chung, J.~Byun, S.~Choe, and M.-S. Choi,
  ``Diffusion-based generative speech source separation,'' \emph{arXiv preprint
  arXiv:2210.17327}, 2022.

\bibitem{moliner2022solving}
E.~Moliner, J.~Lehtinen, and V.~V{\"a}lim{\"a}ki, ``Solving audio inverse
  problems with a diffusion model,'' \emph{arXiv preprint arXiv:2210.15228},
  2022.

\bibitem{chung2022diffusion}
H.~Chung, J.~Kim, M.~T. Mccann, M.~L. Klasky, and J.~C. Ye, ``Diffusion
  posterior sampling for general noisy inverse problems,'' \emph{arXiv preprint
  arXiv:2209.14687}, 2022.

\bibitem{choi2021ilvr}
J.~Choi, S.~Kim, Y.~Jeong, Y.~Gwon, and S.~Yoon, ``Ilvr: Conditioning method
  for denoising diffusion probabilistic models,'' in \emph{Proceedings of the
  IEEE/CVF International Conference on Computer Vision}, 2021, pp.
  14\,367--14\,376.

\bibitem{kong2020diffwave}
Z.~Kong, W.~Ping, J.~Huang, K.~Zhao, and B.~Catanzaro, ``Diffwave: A versatile
  diffusion model for audio synthesis,'' in \emph{International Conference on
  Learning Representations}.

\bibitem{liu2023audioldm}
H.~Liu, Z.~Chen, Y.~Yuan, X.~Mei, X.~Liu, D.~Mandic, W.~Wang, and M.~D.
  Plumbley, ``Audioldm: Text-to-audio generation with latent diffusion
  models,'' \emph{arXiv preprint arXiv:2301.12503}, 2023.

\bibitem{goel2022s}
K.~Goel, A.~Gu, C.~Donahue, and C.~R{\'e}, ``It’s raw! audio generation with
  state-space models,'' in \emph{International Conference on Machine
  Learning}.\hskip 1em plus 0.5em minus 0.4em\relax PMLR, 2022, pp. 7616--7633.

\bibitem{shchekotov22_interspeech}
I.~Shchekotov, P.~K. Andreev, O.~Ivanov, A.~Alanov, and D.~Vetrov, ``{FFC-SE:
  Fast Fourier Convolution for Speech Enhancement},'' in \emph{Proc.
  Interspeech 2022}, 2022, pp. 1188--1192.

\bibitem{kuleshov2017audio}
V.~Kuleshov, S.~Z. Enam, and S.~Ermon, ``Audio super resolution using neural
  networks,'' \emph{arXiv preprint arXiv:1708.00853}, 2017.

\bibitem{andreev2022hifi++}
P.~Andreev, A.~Alanov, O.~Ivanov, and D.~Vetrov, ``Hifi++: a unified framework
  for bandwidth extension and speech enhancement,'' \emph{arXiv preprint
  arXiv:2203.13086}, 2022.

\bibitem{li2020robutrans}
N.~Li, Y.~Liu, Y.~Wu, S.~Liu, S.~Zhao, and M.~Liu, ``Robutrans: A robust
  transformer-based text-to-speech model,'' in \emph{Proceedings of the AAAI
  Conference on Artificial Intelligence}, vol.~34, no.~05, 2020, pp.
  8228--8235.

\bibitem{shen2020non}
J.~Shen, Y.~Jia, M.~Chrzanowski, Y.~Zhang, I.~Elias, H.~Zen, and Y.~Wu,
  ``Non-attentive tacotron: Robust and controllable neural tts synthesis
  including unsupervised duration modeling,'' \emph{arXiv preprint
  arXiv:2010.04301}, 2020.

\bibitem{kong2020hifi}
J.~Kong, J.~Kim, and J.~Bae, ``Hifi-gan: Generative adversarial networks for
  efficient and high fidelity speech synthesis,'' \emph{Advances in Neural
  Information Processing Systems}, vol.~33, pp. 17\,022--17\,033, 2020.

\bibitem{prenger2019waveglow}
R.~Prenger, R.~Valle, and B.~Catanzaro, ``Waveglow: A flow-based generative
  network for speech synthesis,'' in \emph{ICASSP 2019-2019 IEEE International
  Conference on Acoustics, Speech and Signal Processing (ICASSP)}.\hskip 1em
  plus 0.5em minus 0.4em\relax IEEE, 2019, pp. 3617--3621.

\bibitem{yamagishi2019cstr}
J.~Yamagishi, C.~Veaux, K.~MacDonald \emph{et~al.}, ``Cstr vctk corpus: English
  multi-speaker corpus for cstr voice cloning toolkit (version 0.92),'' 2019.

\bibitem{ljspeech17}
K.~Ito and L.~Johnson, ``The lj speech dataset,''
  \url{https://keithito.com/LJ-Speech-Dataset/}, 2017.

\bibitem{NEURIPS2020_wav2vec2}
A.~Baevski, Y.~Zhou, A.~Mohamed, and M.~Auli, ``wav2vec 2.0: A framework for
  self-supervised learning of speech representations,'' in \emph{Advances in
  Neural Information Processing Systems}, vol.~33, 2020, pp. 12\,449--12\,460.

\bibitem{wvmos}
``Wv-mos: Mos score prediction by fine-tuned wav2vec2.0 model,''
  \url{https://github.com/AndreevP/wvmos}, accessed: 2022-01-20.

\bibitem{binkowski2019high}
M.~Bi{\'n}kowski, J.~Donahue, S.~Dieleman, A.~Clark, E.~Elsen, N.~Casagrande,
  L.~C. Cobo, and K.~Simonyan, ``High fidelity speech synthesis with
  adversarial networks,'' in \emph{International Conference on Learning
  Representations}.

\bibitem{jensen2016algorithm}
J.~Jensen and C.~H. Taal, ``An algorithm for predicting the intelligibility of
  speech masked by modulated noise maskers,'' \emph{IEEE/ACM Transactions on
  Audio, Speech, and Language Processing}, vol.~24, no.~11, pp. 2009--2022,
  2016.

\bibitem{le2019sdr}
J.~Le~Roux, S.~Wisdom, H.~Erdogan, and J.~R. Hershey, ``Sdr--half-baked or well
  done?'' in \emph{ICASSP 2019-2019 IEEE International Conference on Acoustics,
  Speech and Signal Processing (ICASSP)}.\hskip 1em plus 0.5em minus
  0.4em\relax IEEE, 2019, pp. 626--630.

\bibitem{nichol2021improved}
A.~Q. Nichol and P.~Dhariwal, ``Improved denoising diffusion probabilistic
  models,'' in \emph{International Conference on Machine Learning}.\hskip 1em
  plus 0.5em minus 0.4em\relax PMLR, 2021, pp. 8162--8171.

\bibitem{hu2018senet}
J.~Hu, L.~Shen, and G.~Sun, ``Squeeze-and-excitation networks,'' 2018.

\bibitem{crash21}
S.~Rouard and G.~Hadjeres, ``Crash: Raw audio score-based generative modeling
  for controllable high-resolution drum sound synthesis,'' in \emph{Music
  Information Retrieval Conf. (ISMIR)}, 2021, pp. 579--585.

\bibitem{borsos2022audiolm}
Z.~Borsos, R.~Marinier, D.~Vincent, E.~Kharitonov, O.~Pietquin, M.~Sharifi,
  O.~Teboul, D.~Grangier, M.~Tagliasacchi, and N.~Zeghidour, ``Audiolm: a
  language modeling approach to audio generation,'' \emph{arXiv preprint
  arXiv:2209.03143}, 2022.

\bibitem{liu2021voicefixer}
H.~Liu, Q.~Kong, Q.~Tian, Y.~Zhao, D.~Wang, C.~Huang, and Y.~Wang,
  ``Voicefixer: Toward general speech restoration with neural vocoder,''
  \emph{arXiv preprint arXiv:2109.13731}, 2021.

\bibitem{birnbaum2019temporal}
S.~Birnbaum, V.~Kuleshov, Z.~Enam, P.~W.~W. Koh, and S.~Ermon, ``Temporal film:
  Capturing long-range sequence dependencies with feature-wise modulations.''
  \emph{Advances in Neural Information Processing Systems}, vol.~32, 2019.

\bibitem{zavivska2022analysis}
P.~Zaviska and P.~Rajmic, ``Analysis social sparsity audio declipper,''
  \emph{arXiv preprint arXiv:2205.10215}, 2022.

\bibitem{zavivska2019proper}
P.~Zaviska, P.~Rajmic, O.~Mokry, and Z.~Pruvsa, ``A proper version of
  synthesis-based sparse audio declipper,'' in \emph{ICASSP 2019-2019 IEEE
  International Conference on Acoustics, Speech and Signal Processing
  (ICASSP)}.\hskip 1em plus 0.5em minus 0.4em\relax IEEE, 2019, pp. 591--595.

\bibitem{griffin1984signal}
D.~Griffin and J.~Lim, ``Signal estimation from modified short-time fourier
  transform,'' \emph{IEEE Transactions on acoustics, speech, and signal
  processing}, vol.~32, no.~2, pp. 236--243, 1984.

\bibitem{luo2019conv}
Y.~Luo and N.~Mesgarani, ``Conv-tasnet: Surpassing ideal time--frequency
  magnitude masking for speech separation,'' \emph{IEEE/ACM transactions on
  audio, speech, and language processing}, vol.~27, no.~8, pp. 1256--1266,
  2019.

\end{thebibliography}

\end{document}